# Transient Thermal and Electrical Characteristics of a Cylindrical LiFeS$_2$ Cell with Equivalent Circuit Model

Khaled I Alsharif[1], Alexander H Pesch[1], Vamsi Borra[1], Pedro Cortes[1], Eric MacDonald[2], Frank X Li[1], Kyosung Choo[1*]

[1] Youngstown State University, OH, 44555
[2] University of Texas at El Paso, TX, 79902

kialsharif@student.ysu.edu ; apesch@ysu.edu ; vborra@ysu.edu ; pcortes@ysu.edu ; emac@utep.edu ; xli@ysu.edu ;
kchoo@ysu.edu*

**Abstract** – This study examines the discharge behaviour of a cylindrical LiFeS$_2$ cell to evaluate the parameters that can be used to predict and estimate the nonlinear dynamic response of a battery. A linear model is developed to simulate the discharge behaviour and examine the thermal behaviour. In particular, a commercial-grade battery is discharged with the industry-standard hybrid power pulsing characterization (HPPC) test and the current and voltage responses are recorded. The dynamic system is modelled with the equivalent circuit model (ECM) through MATLAB Simulink. A block diagram representation of the equivalent circuit model governing equations was developed. The parameter estimation tool was utilized to reduce the error and fit the simulation results to the experimental voltage responses, in order to obtain state of charge dependent dynamic parameters. Those parameters were then used in a Dual-Potential Multi-Scale Multi-Domain (MSMD) Battery Model solved in ANSYS Fluent to analyze the thermal behaviour by acquiring the temperature profiles and the temperature distribution within the cell. The nonlinear behaviour of the battery was characterized and the equivalent circuit model parameters were identified and are shown to agree with the experimental voltage responses. Furthermore, it is found that the battery temperature increased by 7.35°C and was distributed uniformly within the cell.

## 1. Introduction

Electrochemical batteries have been proven to be capable of possessing high energy density, light weight, low-self discharge and other desirable features [1]. Lithium-ion batteries are typically the favourite electrochemical power source to provide high power and long service life [2]. An accurate model of the current and voltage characteristics is critical to optimize in the energy that can be delivered by the battery. This paper investigates the performance of a popular, commercially available cell utilizing the equivalent circuit model [3] and computational fluid dynamics MSMD battery model to identify the dynamic parameters and examine the thermal behaviour of the battery.

Numerous studies have been conducted and developed to establish a better understanding of the nonlinear response behaviour of an electrochemical cell [4-5]. A common technique of characterizing a battery cell is the Electrochemical Impendence Spectrum (EIS), which is employed by applying a small sinusoidal excitation current to the cell at a frequency range of 1 mHz to 10 kHz and measuring its voltage response. The dynamics of the battery is then characterized through Randles equivalent circuit model through the Nyquist and bode plots to identify the internal resistances of the battery [6]. Though valuable insights can be concluded from this technique, however, it cannot predict battery run time and evaluates the dynamic behaviour at a fixed state of charge.

It is highly critical to portray the battery's nonlinear dynamics throughout the entire discharge period; therefore, this study utilizes the hybrid power pulsing characterization (HPPC) test to identify instantaneous and transient parameters of a Lithium-ion battery for the entire state of charge range. An equivalent circuit is then modelled on MATLAB Simulink to describe the battery terminal voltage-current dynamics by utilizing passive electrical circuit components such as capacitors and resistors. To ensure good fitment and accurate representation of the dynamic performance of the battery, the parameter estimation tool is used to fit the simulation results to the voltage response measured experimentally. The parameters obtained from the equivalent circuit model are then employed in the MSMD battery model in ANSYS Fluent to examine the thermal behaviour of the battery.

The next section is a detailing of the battery study conducted and the experimental test setup. Then, the details of the equivalent circuit model are presented and shown to agree with experimental voltage-current data. Next, a CFD model is presented with the thermal performance of the battery. Finally, the paper is concluded with some closing remarks.

## 2. Battery Details and Experimental Quantification

In this study a non-rechargeable Energizer Ultimate Lithium AA battery is considered. The parameters for the battery are displayed in Table 1 [7].

Table 1 : Battery Parameters

| Parameter | Value |
|---|---|
| Name | Energizer Ultimate Lithium |
| Nominal Capacity [mAh] | 3000 |
| Nominal Voltage [V] | 1.5 |
| Cut-off Voltage [V] | 0.8 |
| Form Factor | AA |
| Dimensions [mm] | $14.5 \times 50.5$ |

The hybrid pulse power characterization (HPPC) is a testing method to characterize the dynamic power capability of batteries and the voltage response to step pulses of current at a constant rate. This method has been widely utilized as a technique to identify the required parameters for battery simulations to characterize the nonlinear dynamic behaviour of an electrochemical battery cell [8]. To preform this test, the experimental setup shown in Fig. 1 is employed. This single loop circuit is constructed from a DC power supply (Keysight E36312A) as a source of constant voltage and constant current, a Relay to control the voltage and current signals supplied to the battery in the circuit, a Waveform Generator (Keysight 3360A) to produce square waves triggering the Relay at a set frequency. To measure the voltage and current response of the battery, two Digit Multi-meters (Keysight 34465A 6 1/2) were used, one connected in series with the battery to measure the current and the other connected in parallel across the battery to measure the voltage.

The battery was discharged at a C/2 rate; therefore, the cell was pulsed at 1.5A supplied by the DC power supply. To produce the square waves profile of the current, the waveform generator triggered the relay at a constant frequency of 2.8 mHz with 50% duty cycle. Furthermore, the sampling rate of the Multi-meters were set to 0.4 Sa/s and live data acquisition readings were observed via the BenchVue desktop app. The data recorded was then exported to a MATLAB data file (.mat) for analysis.

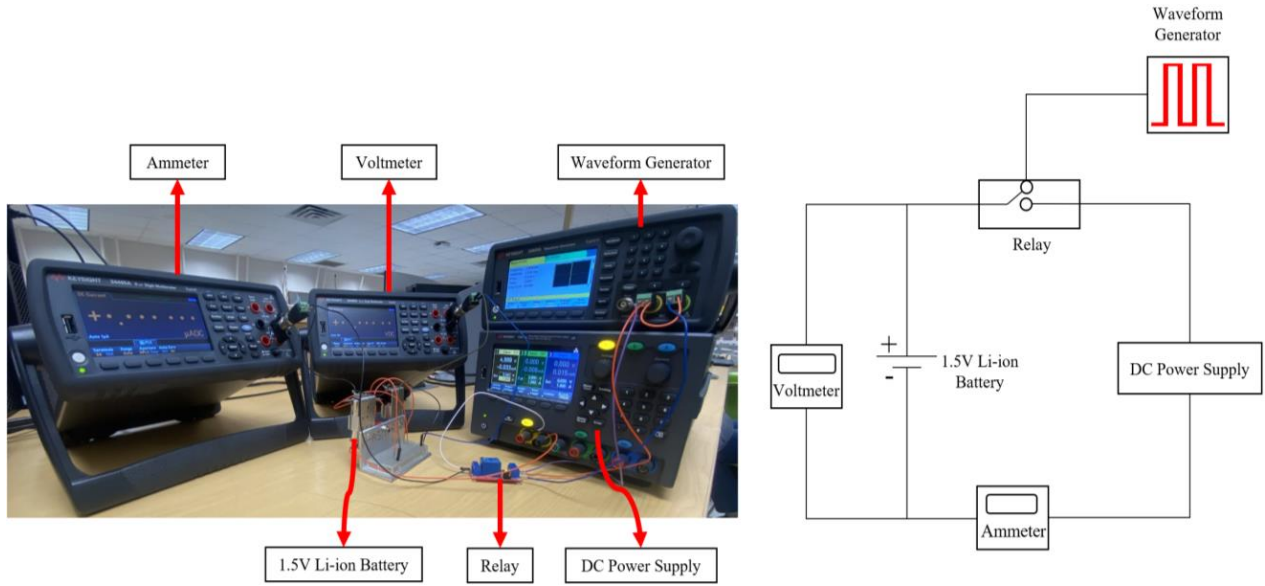

Fig. 1 : Experimental Setup

Displayed in Fig. 2 are the current and voltage responses of the battery measurements obtained from the experiment. The battery is fully discharged as the full range of usable voltage specified by the manufacturer is observed, therefore, the equivalent circuit parameters can be tuned and identified over the fully range of state of charge (SOC).

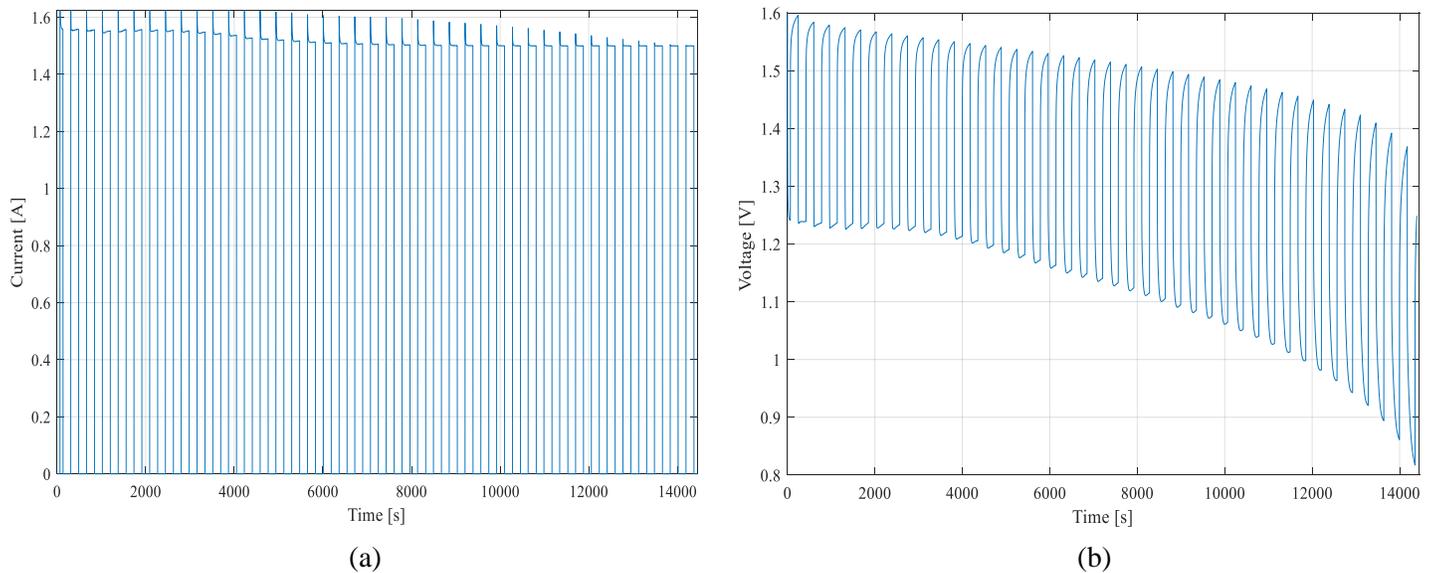

(a)          (b)

Fig. 2 : Experimental results (a) Current measurements (b) Voltage measurements

## 3. Equivalent Circuit Model

The classic equivalent circuit model utilized to characterise the dynamics of the battery is shown in Fig. 3. The voltage source in the circuit represents the open-circuit voltage ($V_{OCV}$), which describes the voltage cross the battery terminals when the cell is unloaded and in complete equilibrium. The resistor in series ($R_s$) is the Ohmic resistance, which accounts for the power dissipated by the battery as heat energy [9]. The two resistor-capacitor sub circuits ($R_1$, $C_1$, $R_2$ and $C_2$) represent the time constants in the circuit which account for the diffusion dynamics of the battery. The Voltage V(*t*) is the measured voltage drop across the passive components of the circuit subtracted from the open-circuit voltage to obtain the losses of the battery. The input to this circuit is the current measurements obtained experimentally and is denoted by I(*t*).

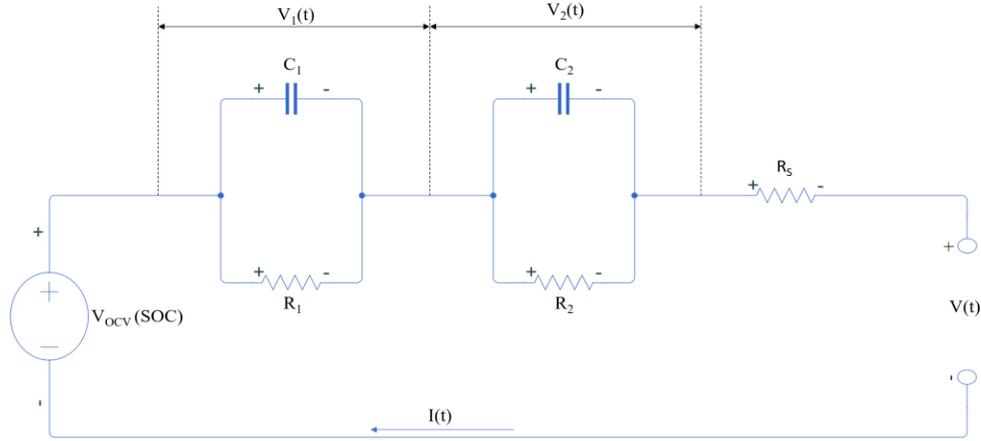

Fig. 3 : Equivalent Circuit Model

The state of charge (SOC) of the battery was solved utilizing Eq. (1). Where $\eta$ is the columbic efficiency or the charge efficiency and is a unitless constant which has a value of 1 when the battery is discharging [9]. $Q$ is the nominal capacity of the battery described in Table 1. Furthermore, the voltage-current relationship was obtained by solving Eqs. (2)-(4).

$$\frac{d(SOC)}{dt} = \frac{-\eta\, I(t)}{Q} \tag{1}$$

$$V = V_{OCV}(SOC) - V_1 - V_2 - R_S(SOC)I(t) \tag{2}$$

$$\frac{dV_1}{dt} = -\frac{1}{R_1(SOC)C_1(SOC)}V_1 - \frac{1}{C_1(SOC)}I(t) \tag{3}$$

$$\frac{dV_2}{dt} = -\frac{1}{R_2(SOC)C_2(SOC)}V_2 - \frac{1}{C_2(SOC)}I(t) \tag{4}$$

To solve the equations described above, the block diagram shown in Fig. 4 was employed utilizing MATLAB Simulink. The $V_{OCV}$ was identified from the peaks of the voltage response during the relaxation period and the passive circuitry components were modelled as 1-D lookup tables with 20 different values that correspond to the state of charge at different time intervals. The parameter estimator toolbox in Simulink was utilized to reduce the error between the simulation and experimental voltage measurements, therefore, the capacitor and the resistor values in the lookup tables were tuned to fit the simulation voltage to the experimental voltage.

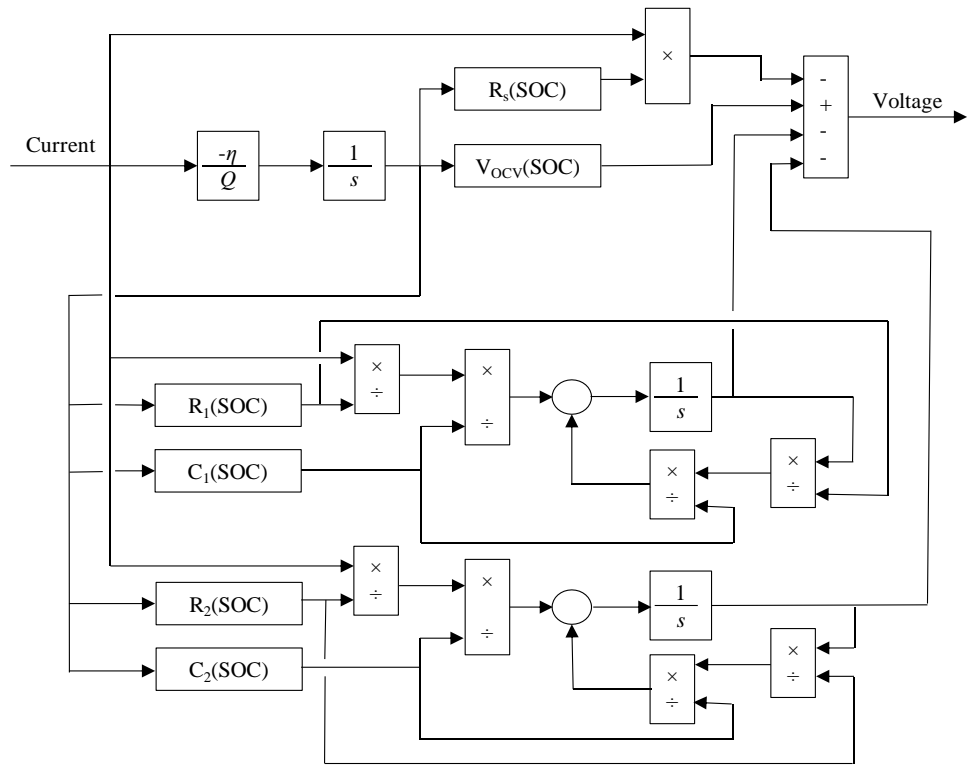

Fig. 4 : Equivalent circuit model block diagram in MATLAB Simulink

The results obtained from the Simulink simulation for the state of charge degradation over time, the open-circuit voltage as a function of the State of Charge and the measured and simulated voltage after utilizing the parameter estimation toolbox are displayed in Fig. 5.

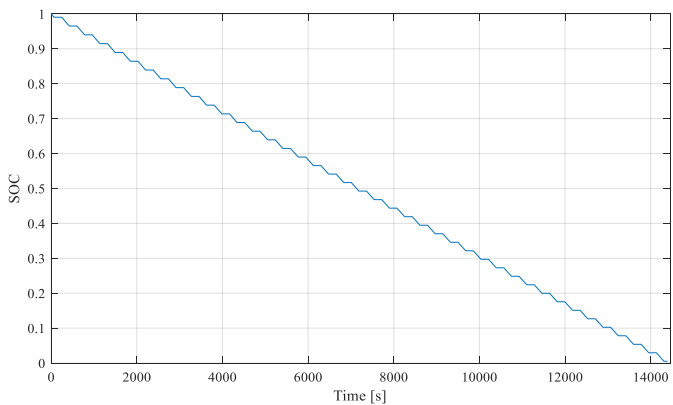

(a)

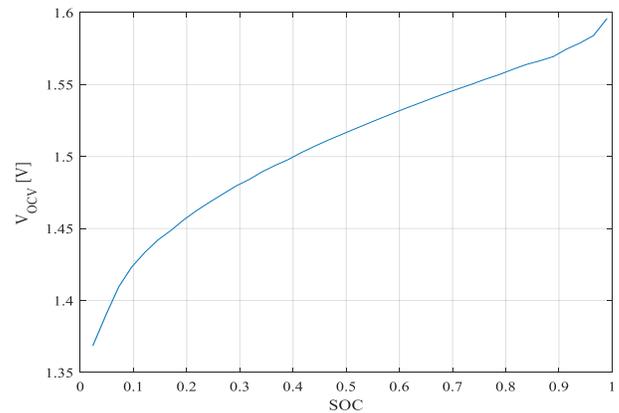

(b)

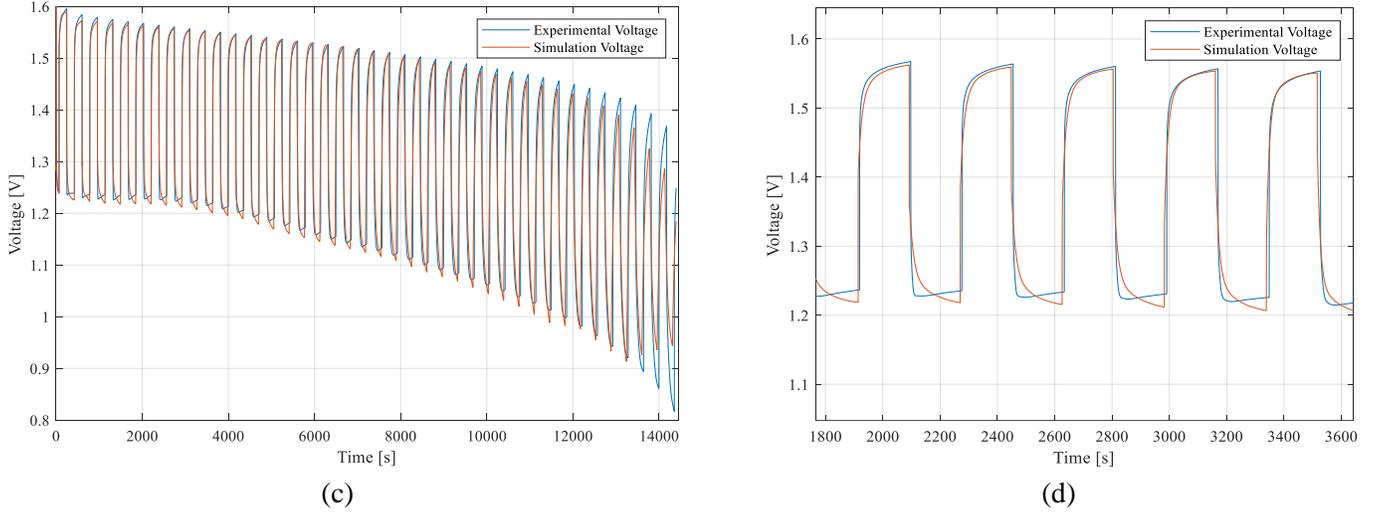

Fig. 5 : (a) SOC per coulomb counting (b) OCV as a function of SOC (c) Experimental voltage response and model voltage after parameter estimation (d) Close-up illustration of the simulated results and the experimental measurements

As illustrated in Fig. 5(a), the battery is fully discharged through the continuous constant pulses. Additionally, the Open-Circuit Voltage shown in Fig.5(b), was characterized through a 5$^{th}$ order polynomial equation as a function of SOC with an R$^2$ value of 0.9995 as shown in Eq. (5).

$$V_{OCV}(SOC) = 2.33(SOC)^5 - 6.36(SOC)^4 + 6.62(SOC)^3 - 3.35(SOC)^2 + (SOC) + 1.35 \tag{5}$$

Furthermore, Fig. 5(c) displays the agreement of the Simulink simulation voltage after utilizing the parameter estimation tool to the battery response voltage measured experimentally. Fig. 5(d) shows a closeup illustration of the voltage results for the pulses that occurred in the interval of $1800s \leq t \leq 3600s$ to express how well the simulated parameters were fitted onto the measured data. The simulated results are within range to be deemed accurate to obtain the state of charge dependent dynamic variables. The tuned values of the passive components in the circuit are demonstrated in Table 2.

Table 2 : Estimated parameters as a function of state of charge

| SOC | Rs (Ω) | R1 (Ω) | R2 (Ω) | C1 (F) | C2 (F) |
|---|---|---|---|---|---|
| 0.05 | 2.60E-05 | 2.68E-01 | 8.51E-02 | 2.33E+02 | 4.00E+03 |
| 0.10 | 2.32E-04 | 3.47E-01 | 1.71E-03 | 1.92E+02 | 3.08E+03 |
| 0.15 | 2.66E-02 | 2.98E-01 | 7.76E-04 | 1.99E+02 | 5.33E+03 |
| 0.20 | 1.13E-02 | 3.12E-01 | 3.44E-04 | 1.73E+02 | 3.70E+03 |
| 0.25 | 1.07E-02 | 2.82E-01 | 1.55E-02 | 1.82E+02 | 3.89E+03 |
| 0.30 | 1.48E-02 | 2.67E-01 | 1.45E-02 | 1.85E+02 | 1.28E+03 |
| 0.35 | 1.18E-02 | 2.64E-01 | 8.72E-03 | 1.76E+02 | 1.23E+03 |
| 0.40 | 1.29E-02 | 2.56E-01 | 9.20E-03 | 1.68E+02 | 1.34E+03 |
| 0.45 | 2.14E-02 | 2.39E-01 | 6.93E-03 | 1.71E+02 | 1.39E+03 |
| 0.50 | 2.82E-02 | 2.22E-01 | 9.52E-03 | 1.60E+02 | 1.54E+03 |
| 0.55 | 3.19E-02 | 2.01E-01 | 2.11E-02 | 1.21E+02 | 1.65E+03 |
| 0.60 | 3.19E-02 | 1.91E-01 | 2.47E-02 | 1.05E+02 | 1.78E+03 |
| 0.65 | 3.25E-02 | 1.80E-01 | 2.89E-02 | 8.71E+01 | 1.86E+03 |
| 0.70 | 3.13E-02 | 1.68E-01 | 3.26E-02 | 1.74E+01 | 1.91E+03 |
| 0.75 | 5.49E-02 | 1.40E-01 | 3.16E-02 | 6.40E+01 | 1.86E+03 |
| 0.80 | 8.20E-02 | 1.09E-01 | 3.27E-02 | 1.09E+02 | 2.01E+03 |
| 0.85 | 1.11E-01 | 8.03E-02 | 3.27E-02 | 1.62E+02 | 2.08E+03 |
| 0.90 | 1.46E-01 | 5.24E-02 | 3.20E-02 | 2.22E+02 | 2.46E+03 |
| 0.95 | 1.64E-01 | 4.05E-02 | 3.25E-02 | 2.77E+02 | 3.50E+03 |
| 1.00 | 1.75E-01 | 5.45E-02 | 6.80E-02 | 4.20E+02 | 9.24E+03 |

## 4. Computational Fluid Dynamics

The ANSYS Fluent Dual Potential Multi-Scale Multi-Domain (MSMD) battery model was used to determine the temperature profile and distribution at the battery's length scale. To obtain the voltage drop across the battery, fluent establishes the current-voltage relationship from Eqs. (1) – (4). The Li-ion transport occurs in the anode-separator-cathode sandwich layers [10]. Therefore, Eqs. (6) – (10) are the governing equations solved in the CFD domain at the battery cell's scale, which describe the thermal and electric fields, respectively.

$$\frac{\partial \rho C_p T}{\partial t} - \nabla \cdot (k \nabla T) = \sigma_+ |\nabla \varphi_+|^2 + \sigma_- |\nabla \varphi_-|^2 + \dot{q}_{ECh} \tag{6}$$

$$\nabla \cdot (\sigma_+ \nabla \varphi_+) = -j_{ECh} \tag{7}$$

$$\nabla \cdot (\sigma_- \nabla \varphi_-) = j_{ECh} \tag{8}$$

$$j_{ECh} = I \frac{Q}{Q_{ref} Vol} \tag{9}$$

$$\dot{q}_{ECh} = j_{ECh} [V_{OCV} - V - T \frac{dU}{dT}] \tag{10}$$

Where $\sigma_+$ and $\sigma_-$ and $\varphi_+$ and $\varphi_+$ are the electric conductivities and the phase potentials for the positive and negative electrodes, correspondingly. Additionally, $j_{ECh}$ and $\dot{q}_{ECh}$ are the volumetric current transfer rate and the electrochemical reaction heat due to electrochemical reactions [10].

Fig. 6 illustrates the geometry of the battery modelled and the mesh. The geometry of the Active zone of the battery cell was modelled according to the dimensions provided by Energizer [7]. A hex-dominate quadratic mesh chosen with an element size of 0.19 mm which yielded to a total of 946964 elements.

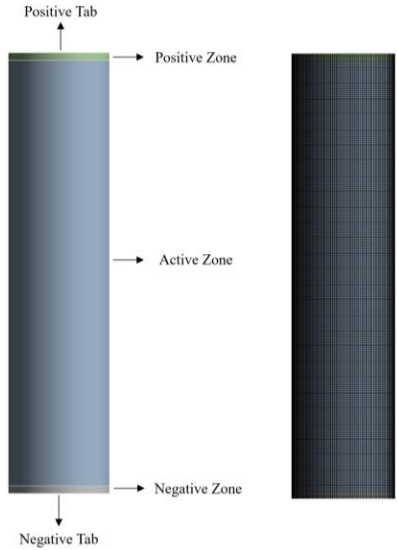

Fig. 6 : Geometry definition and model mesh

The active zone was chosen as the active component and the positive and negative zones were selected as the passive components of the cell. Furthermore, the positive and negative tab were considered as the external connections of the battery.

The open-circuit voltage attained from Eq. (5) and the passive electrical components values listed in Table 2, were utilized as the model parameters for this simulation.

Fig. 7 displays the temperature profile and the temperature distribution in the battery at a discharge rate of C/2. Fig. 7(a) presents the temperature profile throughout the discharge period and demonstrates an overall temperature increase of 7.35 °C. From the contours displayed in Fig. 7(b), it can be observed that the temperature distribution within the cell is uniform.

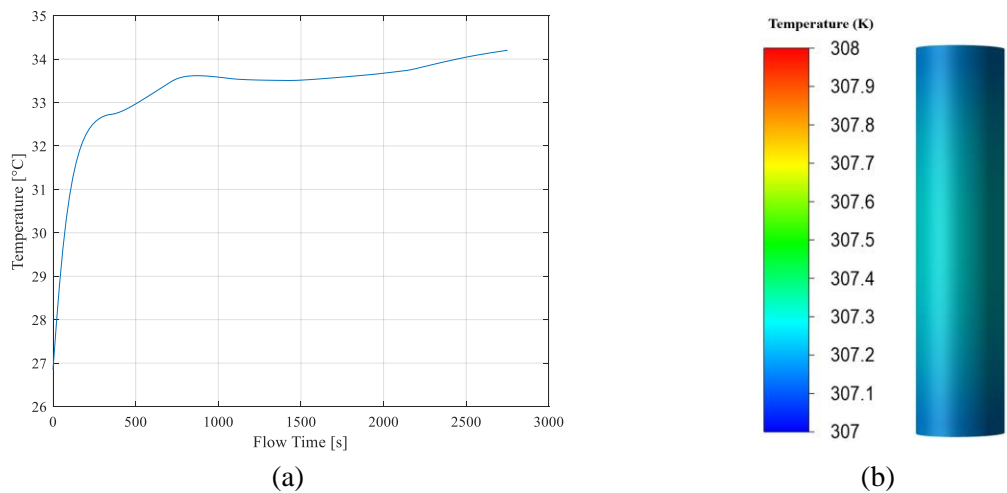

(a)      (b)

Fig. 7 : CFD results (a) Temperature profile (b) Temperature contour

## 5. Conclusion

The study conducted in this paper investigated a linear model to characterize the dynamic and thermal behaviour of a cylindrical LiFeS$_2$ battery cell. The dynamic parameters of the battery were modelled with the equivalent circuit model utilizing passive circuit components. A block representation of the governing equations of the equivalent circuit was employed and simulated on MATLAB Simulink and the results were fitted to the experimental voltage response using the parameter estimation toolbox to obtain state of charge dependent variables. Furthermore, these variables were used as the model parameters to the MSMD battery model in ANSYS Fluent to predict the thermal performance of the battery.

The open-circuit voltage was evaluated from the peaks of the voltage relaxation period and characterized with a 5$^{th}$ degree polynomial equation. Additionally, the simulated voltage was shown to be in good agreement with the experimental voltage after utilizing the parameter estimation toolbox, as a result, the passive components of the circuit were obtained at different state of charge time intervals for the full discharge period. The CFD simulation shows a 7.35°C temperature increase and a uniform temperature distribution within the cell.

**Acknowledgements**

… ADMETE…